\documentclass[aps,prl,reprint,groupedaddress,amsfonts,amssymb,amsmath]{revtex4-1}

\usepackage{graphicx}
\usepackage{dcolumn}
\usepackage{bm}
\usepackage[mathlines]{lineno}
\usepackage{siunitx}
\usepackage{booktabs}
\usepackage{color}
\usepackage{multirow}
\usepackage{lipsum}

\begin{document}

\title{Gap-opening transition in Dirac semimetal ZrTe$_5$}

\author{Yefan Tian}
 \email{yefantian93@gmail.com}
\affiliation{Department of Physics and Astronomy, Texas A\&M University, College Station, TX 77843, USA}
\author{Nader Ghassemi}
\affiliation{Department of Physics and Astronomy, Texas A\&M University, College Station, TX 77843, USA}
\author{Joseph H. Ross, Jr.}
\email{jhross@tamu.edu}
\affiliation{Department of Physics and Astronomy, Texas A\&M University, College Station, TX 77843, USA}

\date{\today}

\begin{abstract}
We apply $^{125}$Te nuclear magnetic resonance (NMR) spectroscopy to investigate the Dirac semimetal ZrTe$_5$. With the NMR magnetic field parallel to the $b$-axis, we observe significant quantum magnetic effects. These include an abrupt drop at 150 K in spin-lattice relaxation rate. This corresponds to a gap-opening transition in the Dirac carriers, likely indicating the onset of excitonic pairing. Below 50 K, we see a more negative shift for the Te$_z$ bridging site indicating the repopulation of Dirac levels with spin polarized carriers at these temperatures. This is the previously reported 3D quantum Hall regime; however, we see no sign of a charge density wave as has been proposed.
\end{abstract}

\maketitle


ZrTe$_5$ has recently been widely studied due to its exotic electronic properties and topological nature. Although initially of interest because of a resistance anomaly which was proposed to indicate a charge density wave (CDW), no evidence was found for a CDW in zero field \cite{okada1982negative}. Nevertheless, in an applied magnetic field, Dirac materials can be strongly susceptible to formation of phases such as density waves, Axion insulators, or nematic phases \cite{wei2012excitonic,wang2013chiral,roy2015magnetic,liu2016zeeman,zhang2016topological}. ZrTe$_5$ specifically has shown remarkable behavior such as the chiral magnetic effect \cite{li2016chiral} and 3D quantum Hall effect (3DQHE) \cite{liu2016zeeman,tang2019three} in a magnetic field. Anomalous thermoelectric effects are also observed in the quantum limit \cite{zhang2019anomalous}, possibly connected to the unusual dispersion behavior \cite{martino2019two,jiang2020unraveling}. 

Excitonic insulators may also be induced by electronic interactions \cite{jerome1967excitonic}. In Dirac systems, particle-hole symmetry can promote formation of electrons and holes bound by the Coulomb force. The resulting condensate generates a finite energy gap at the Dirac point and turns the semimetal into an excitonic insulator \cite{kotov2012electron}. There has been much recent interest in systems which may form such a state, and in nodal-line Dirac semimetals ZrSiS and ZrSiSe \cite{scherer2018excitonic,rudenko2018excitonic,wang2020quantum}, it has been proposed that the enhanced density of Dirac states in the vicinity of the node may promote such a ground state. 

\begin{figure}
\includegraphics[width=\columnwidth]{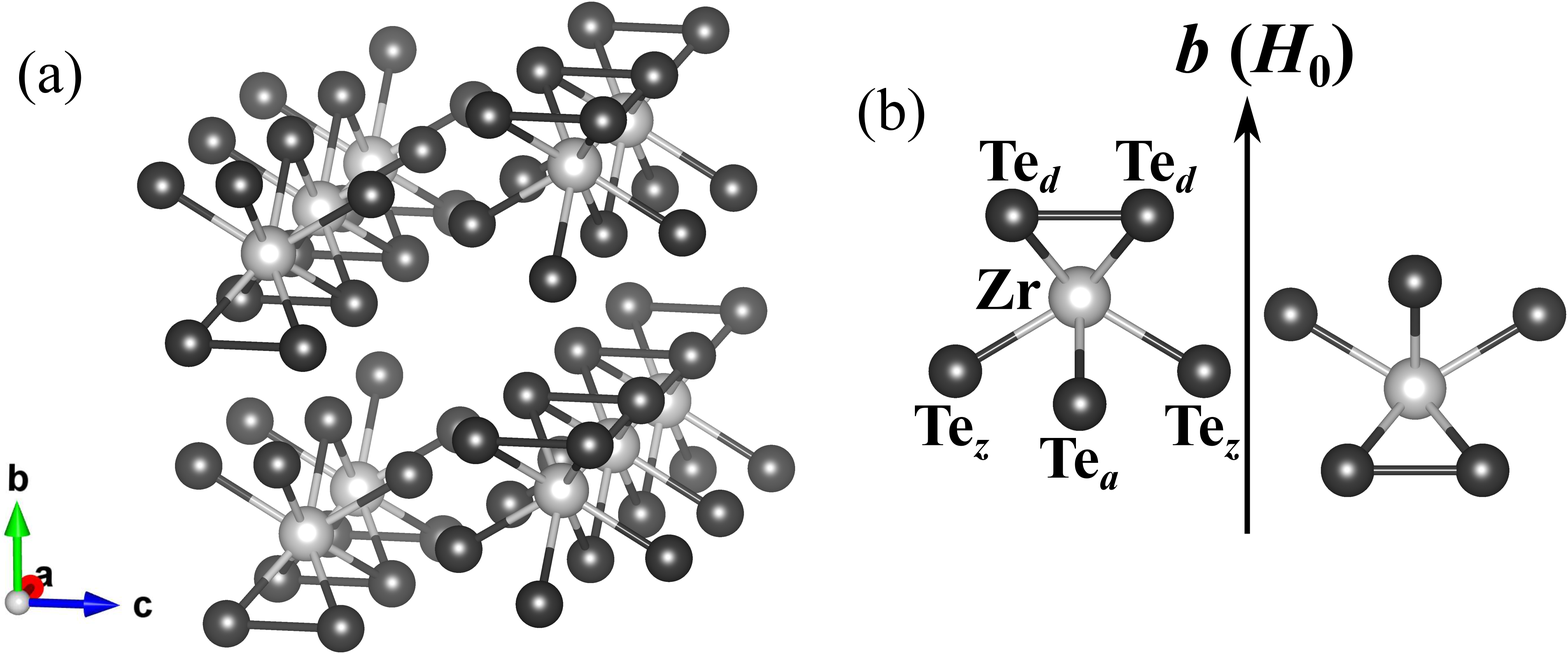}
\caption{\label{structure} (a) Layered structure of ZrTe$_5$. (b) View along $a$ with Te$_d$, Te$_z$, and Te$_a$ labeled, showing the NMR field ($H_0$) along $b$.}
\end{figure}

In this Letter, we examine magnetic quantum effects in ZrTe$_5$ using $^{125}$Te NMR with magnetic field parallel to $b$ (Fig.~\ref{structure}). Among the results discussed, we find that at 150 K, a clear change in spin-lattice relaxation time results corresponding to gap opening indicating possible exciton states. At low temperatures, we see no evidence for a field-induced CDW.

\begin{figure*}
\includegraphics[width=2\columnwidth]{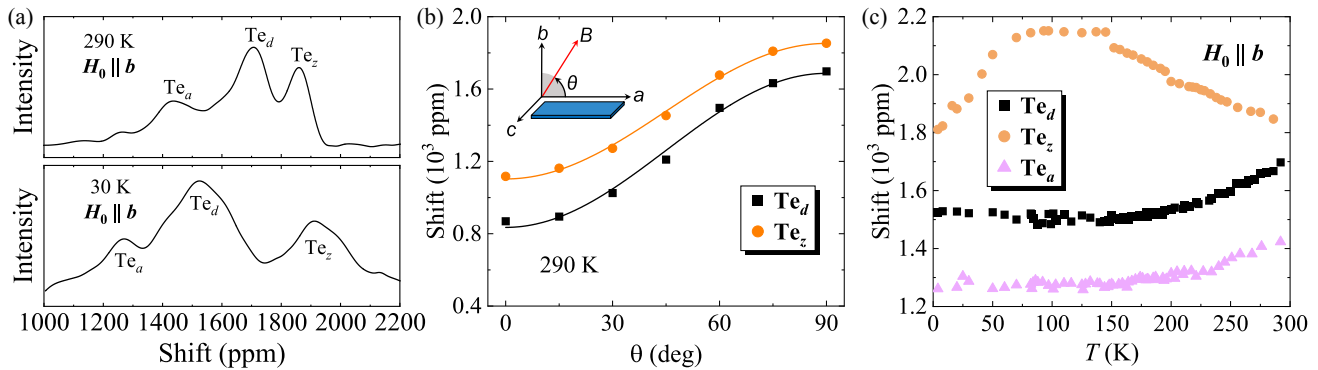}
\caption{\label{shift}(a) $^{125}$Te Lineshapes at 290 K and 30 K labeled for 3 Te sites. (b) Angular dependence at 290 K. (Te$_a$ not mapped at all angles due to low signal strength for this transition). Solid curves: fits to $K \propto A+B\cos^2\theta$. (c) $H_0 \parallel b$ shift vs $T$.}
\end{figure*}

The NMR sample containing many aligned crystals from a single chemical vapor transport (CVT) growth batch was described in Ref.~\cite{tian2019dirac}. Comparing the 125 K resistivity peak for a representative crystal \cite{tian2019dirac} to reported 95 and 135 K positions for samples with $n = 1.4$ and $10 \times 10^{17}$ cm$^{-3}$ \cite{tang2019three,shahi2018bipolar}, we estimate for our crystals $n = 5 \times 10^{17}$ cm$^{-3}$. NMR experiments utilized a custom-built spectrometer at a fixed field $H_0\approx9$ T. $^{125}$Te shifts were calibrated by aqueous Te(OH)$_6$ and adjusted for its $\delta=707$ ppm paramagnetic shift to the dimethyltelluride standard \cite{inamo1996125te}.



Fig.~\ref{shift}(a) shows the $^{125}$Te lineshapes at 290 K and 30 K with $H_0 \parallel b$ for which orientation the field has a particularly large effect on the Dirac carriers \cite{shahi2018bipolar}. The three peaks correspond to the three Te sites (Fig.~\ref{structure}). Shift positions are identified as the fitted maximum intensity positions. Fig.~\ref{shift}(b) depicts the dependence on the angle between the $a$-axis and the field $H_0$, with a $\cos^2\theta$ fit as expected for linear response to the field at this temperature. The previous assignment \cite{tian2019dirac} for $H_0 \parallel a$ yields the site identities shown in Fig.~\ref{shift}(a).

\begin{figure*}
\includegraphics[width=2\columnwidth]{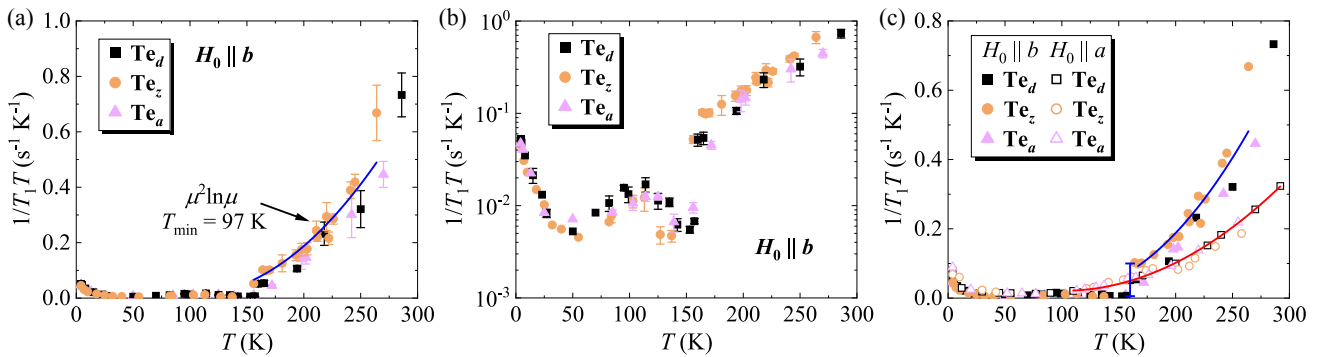}
\caption{\label{T1TvsT} $1/T_1T$ vs $T$ in (a) linear scale and (b) log scale. (c) Comparison between $H_0 \parallel b$ and $H_0 \parallel a$ data from Ref.~\cite{tian2019dirac}. Solid curves: $\mu^2\ln\mu$ fits described in text.}
\end{figure*}

Spin-lattice relaxation was measured by inversion recovery and well-fitted to $M(t)=M(\infty)(1-Ce^{-t/T_1})$, yielding results shown in Fig.~\ref{T1TvsT}. With ZrTe$_5$ Dirac and band-edge states dominated by Te $p$-states \cite{weng2014transition}, core polarization and dipolar hyperfine coupling would be expected to play significant roles. In most cases, these terms cause site dependence. Instead, the behavior shown in Fig.~\ref{T1TvsT} is independent of site for the whole temperature range, similar to the results for $H_0 \parallel a$ \cite{tian2019dirac}. 

At high temperatures, $1/T_1T$ follows a $\mu^2\ln\mu$ curve [Fig.~\ref{T1TvsT}(a)], assuming $\mu \propto T$ with a $T$-independent background \cite{tian2019dirac}. Fig.~\ref{T1TvsT}(c) compares $H_0 \parallel a$ \cite{tian2019dirac} and $H_0 \parallel b$. The $\mu^2\ln\mu$ behavior is similar for both orientations, due to a long-range orbital mechanism for 3D Dirac systems \cite{okvatovity2019nuclear,maebashi2019nuclear}, which can operate when high mobility carriers are present. For $H_0 \parallel b$, the results extrapolate to a crossing of the Dirac node by the chemical potential at $T_0=97$ K. Compared to $T_0=85$ K \cite{tian2019dirac} for $H_0 \parallel a$, this is consistent with the field-induced increase of the resistance-anomaly temperature which is absent for $H_0 \parallel a$ \cite{shahi2018bipolar}. 

At 150 K, there is a sudden $1/T_1T$ drop [Fig.~\ref{T1TvsT}(b)]. This is an indication of a reduction of $g(E_F)$, since the extended orbital $T_1$ mechanism involves states within about $k_BT$ of $E_F$, as is the general case for relaxation induced by charge carriers. Thus, this change must correspond to a gap opening in the high-mobility Dirac carriers associated with the $\mu^2\ln\mu$ behavior for $T > 150$ K. 

Near 120 K, there is a small $1/T_1T$ peak [Fig.~\ref{T1TvsT}(b)]. This is consistent in temperature and magnitude with what has been observed in other low-carrier density materials \cite{tian2018native}, caused by dipolar coupling to dilute paramagnetic moments, such as perhaps the donors causing $n$-type behavior here. Similar to the long-range orbital mechanism, this term will be independent of site. Since this contribution and any remaining charge carrier terms will be additive, it can be seen that the reduction in carrier-based $1/T_1T$ is at least an order of magnitude.


Fig.~\ref{shift}(c) shows the $T$-dependence of the shifts. The bridging Te$_z$ site exhibits distinctive behavior, as was also the case for $H_0 \parallel a$ \cite{tian2019dirac}. In the present case, distinctive features are associated with the quantum magnetic response for this orientation. The shifts can generally be divided into a Knight shift ($K$) due to the spin response of carriers and chemical shift associated with the orbital susceptibility. For Dirac systems, the orbital term can include an enhanced diamagnetic response \cite{okvatovity2019nuclear,maebashi2019nuclear,tian2020topological,tay2020unusual,wang2020landau}; however, this term is likely to be small for the present 3D Dirac case, and we do not observe the characteristic temperature dependence. Other than such a term, chemical shifts are typically slowly varying with $T$, and likely constant at low $T$ as seen in the Te$_a$ and Te$_d$ data. The $T$-dependence for the Te$_z$ is shown below to be consistent with a core-polarization-induced Knight shift, and thus for convenience we denote the entire shift as $K$.


DFT calculations \cite{tian2019dirac} indicate that aside from the Dirac crossing at $\Gamma$, there is also a conduction band (CB) minimum along the $Y$-$X_1$ direction, approximately 20 meV above the Dirac node. Thus, we model this system in a 3-band approximation, including the Dirac electrons and holes, and quadratic CB pocket. For the Dirac-cone dispersion, magneto-optic measurements \cite{martino2019two} are consistent with approximately quadratic dispersion along $k_b$, and linear in other directions, while other recent work \cite{morice2020optical,jiang2020unraveling} also points to a non-Dirac dispersion along $k_b$. To include the effect of a flatter $k_b$ dispersion, we compare the limiting cases of a 3D linear Dirac cone, and the quasi-2D case with no dispersion along $k_b$.

\begin{figure*}
\includegraphics[width=2\columnwidth]{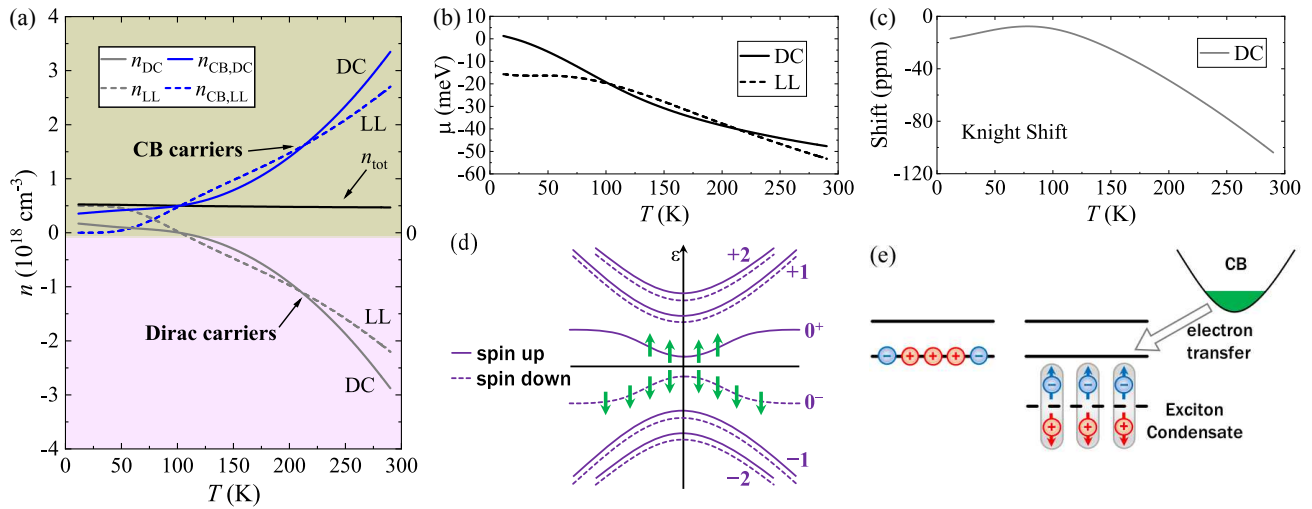}
\caption{\label{discussion} (a)-(c) Numerical results vs $T$ for continuous Dirac (DC) and quasi-2D Landau level (LL) models in text: (a) Carrier concentrations. (b) Chemical potential. (c) Knight shift. (d) Sketch of Landau levels and low-$T$ spins. (e) Condensation process with CB pocket participation.}
\end{figure*}

Results from such a model are shown in Figs.~\ref{discussion}(a)-(c), for a 3D Dirac cone with $v_F = 2.1 \times 10^5$ m/s, from the product of the 3 principal Fermi velocities reported in Ref.~\cite{tang2019three}, and $n = 5 \times 10^{17}$ cm$^{-3}$ assumed fixed due to native doping. We assume an effective mass $m^* = m_e$ for the quadratic CB, while the Dirac cone with linear dispersion has a density of states (both spins) $g(E) = E^2/[\pi^2 (\hbar v_F )^3]$. We also assign a $g$-factor of 22.5 for the Dirac states, a mean of the reported results \cite{chen2015magnetoinfrared,liu2016zeeman,sun2020large}. Solving numerically with $n$ fixed, we obtain the $T$-dependent CB and Dirac carrier densities, and corresponding chemical potential ($\mu$) shown in Fig.~\ref{discussion} (curves labeled DC). Near room temperature the CB electron and Dirac hole densities become large, a result which is not very sensitive to the total $n$. The quasi-2D Dirac case gives qualitatively similar results, see Supplemental Material \cite{sm}. In the results, $\mu$ crosses the Dirac node near 100 K, in agreement with the fitting of the high-$T$ $1/T_1T$ giving an extrapolated node crossing of 97 K.

The large Dirac $g$-factor also leads to an increasingly more negative $K$ as room temperature is approached in this model. This was calculated from the spin density difference of Dirac cones shifted by the Zeeman energy $\pm g\mu_B\mu_0H_0/2$ in the 9 T NMR field. Assuming the core polarization hyperfine interaction dominates for Te $p$-states with $H_\mathrm{HF} = –15$ T$/\mu_B$ \cite{carter1977metallic}, and the spin density locates on the Te$_z$ sites, $K$ is the ratio of the net hyperfine field to $H_0$. The calculated trend shown in Fig.~\ref{discussion}(c) agrees with the measured Te$_z$ shift above 150 K, although the magnitude is about 3 times smaller than observed -- previous NMR measurements for $H_0 \parallel a$ \cite{tian2019dirac} indicated a somewhat faster change of $\mu$ vs $T$ than given in this model, which may be due to excitation of carriers into additional CB pockets which exist at higher energies.

The model described above assumed continuous Dirac states; however, at low $T$ discrete Landau levels will be important, as sketched in Fig.~\ref{discussion}(d). Landau Level separations \cite{burkov2011topological} at $k_b = 0$ are $E_N = \mathrm{sgn}(N)\sqrt{2e\hbar\mu_0 H_0 v_F^2 |N|}$, with each single-spin level occupied by $n_0 = \mu_0 H_0/(\Phi_0 b)$ per volume, where $\Phi_0$ is the flux quantum and $b$ the lattice constant. With $v_F = 5.6 \times 10^5$ m/s (the mean $a$ and $c$-direction values \cite{martino2019two}), we obtain the separation between $N = 0$ and $\pm1$ levels $E_1 = 61$ meV, and $n_0 = 1.6\times 10^{18}$ cm$^{-3}$. For $n = 0.5 \times 10^{18}$ cm$^{-3}$, the $N = 0^+$ level is thus about 0.3 filled at $T = 0$. The 9 T Zeeman splitting of the $N = 0$ levels is 10 meV, which may be enhanced by the $T$-dependent Dirac gap \cite{tian2019dirac}. From the Hamiltonian parameters quoted in Ref.~\cite{chen2015magnetoinfrared}, we calculate an energy dispersion of about 10 meV over the range of the filled $N = 0^+$ states, and the proposed flatter $k_b$ dispersion \cite{martino2019two,morice2020optical,jiang2020unraveling} implies a situation closer to quasi-2D discrete Landau levels. Thus, our sample is within the quantum limit in the 9 T measuring field, with only the $N = 0^+$ level occupied at low temperatures in absence of interactions.

Figs.~\ref{discussion}(a)-\ref{discussion}(c) shows results according to the above parameters (labeled LL) with the Landau states treated as quasi-2D. In this case, below about 50 K, the CB becomes depleted in favor of the large density of states in the $0^+$ level. At high temperatures, the overlap of higher levels becomes important and the 3D continuous model better represents the situation. 

The abrupt $1/T_1T$ drop at 150 K indicates the disappearance of the long-range orbital relaxation and hence the depletion of Dirac states near $\mu$, signaling a gap-opening process. There have been many discussions of interactions on Dirac electron systems, with a magnetic field effectively reducing the dimensionality to promote this, and in ZrTe$_5$ the quadratic dispersion along $k_b$ may also enhance these effects \cite{janssen2016excitonic,wang2017excitonic}. Possible resulting states include excitonic condensates, charge or spin density waves, or other symmetry breaking phases \cite{wei2012excitonic,zhang2016topological,liu2016zeeman,pan2019ground}. It seems unlikely that the development of a density wave is responsible since there will be no well-defined Fermi surface at 150 K with $\mu$ approaching the node and the Dirac carrier density undergoing significant changes vs $T$ (Fig.~\ref{discussion}). On the other hand, this situation, with balanced Dirac electron and hole numbers, is favorable for spontaneous exciton formation \cite{jerome1967excitonic}, enhanced by the Dirac electron-hole symmetry. 
 
While exciton condensate formation normally would be continuous, the reservoir of CB carriers provides a likely explanation for the abrupt change, as illustrated in Fig.~\ref{discussion}(e): With holes and electrons nearly equal, transfer of electrons from the CB to attain balance could bootstrap the process since they will aid in the lowering of the condensate energy. This requires a relatively small electron transfer, and not the disappearance of all carriers -- with the exciton condensate consisting of balanced electron and hole numbers, the CB will retain the native carrier density as $\mu$ continues to change vs $T$. The Knight shift results match this scenario, since the large $g$-factor makes $K$ particularly sensitive to the Dirac spins, and the small increase in Te$_z$ shift at 150 K indicates a decrease in spin density due to the negative core polarization hyperfine field $H_\mathrm{HF}$. The constant shift below this corresponds to a zero net-spin configuration, and thus a singlet condensate with no net Dirac spin polarization.

Starting at $\sim$50 K, the Te$_z$ shift begins to decrease [Fig.~\ref{shift}(c)], and by 4 K its magnitude is reduced by almost 200 ppm. This corresponds to an increase of paramagnetic spin density on Te$_z$, because of the negative $H_\mathrm{HF}$ for Te $p$-orbitals. At these temperatures, $\mu$ is expected to be close to the $0^+$ level, with the $n = 0.5 \times 10^{18}$ cm$^{-3}$ available carriers settled on this level [Fig.~\ref{discussion}(a)-\ref{discussion}(b)], and the $0^-$ level completely occupied (by $1.6 \times 10^{18}$ cm$^{-3}$ as noted above). To the extent that spin-orbit coupling does not mix spin configurations, due to the positive $g$-factor the $0^-$ level will contain the paramagnetic-sign spins. This implies a net spin density of $1.1 \times 10^{18}$ cm$^{-3}$. With 8 Te$_z$ sites per 795 \si{\angstrom}$^3$ unit cell \cite{fjellvaag1986structural} this is $1.1 \times 10^{-4}$ spins per Te$_z$ site, and with $H_\mathrm{HF} =-15$ T$/\mu_B$, the net $-1.6$ mT hyperfine field yields an expected $K = (-1.6$ mT$) / (9$ T$) = -180$ ppm, very close to what is measured. 

This scenario implies the dissolution of the exciton condensate which we identify below 150 K. The carriers which settle into Dirac states at low $T$ can lead to such a condensate destabilization \cite{jerome1967excitonic}. An alternative situation in which electrons are simply added to the $0^+$ level would instead cause a positive shift change, the opposite of what we observe below 50 K. The increase in $1/T_1T$ as $T$ is lowered in this regime is also consistent with the reappearance of high-mobility Dirac carriers, based on the extended-orbital mechanism described above, which is expected to operate for the Dirac electrons and extend to all sites as observed. 

Ref.~\cite{tang2019three} proposed a phase diagram which aligns with the quantum effects observed here: At the 150 K boundary between hole- and electron-dominated phases we show that there is a gap opening in the Dirac states. Below 50 K is a series of quantum states; in our sample $n$ is about 4 times larger than Ref.~\cite{tang2019three}, so our results are comparable to those for $B = 9$ T$/4 \sim 2$ T, just above the quantum limit identified, in the 3DQHE regime. 

The quantum limit identification in Ref.~\cite{tang2019three} includes a 4-fold enhancement in areal carrier density associated with a proposed field-induced CDW. Without this enhancement, our sample will be well-past the quantum limit, with level filling of $\sim$0.3 as determined above. We do not find evidence for a CDW in our sample, which normally would cause characteristic NMR splitting \cite{ross1992nuclear}. We do see a low-$T$ increased broadening [Fig.~\ref{shift}(a) inset]; however, to within uncertainty this is a dynamical linewidth, and matches a decrease in the $T_2$ coherence decay time we observe at low temperatures. Because of this enhanced natural width, it could be possible for CDW splitting to be hidden; however, the amplitude must be small: The proposed \cite{tang2019three} 4-unit-cell CDW, if its effect is to confine the Dirac electrons into a conducting layer every 4 cells and thereby promote the 3DQHE, implies a charge density in these layers 4 times larger than for the sample as a whole. Repeating the above estimate of $K = 180$ ppm, this leads to an enhanced $K$ in the conducting layers of about 600 ppm based on the filling factor of our sample. The majority of the NMR line would have no Knight shift, with a net 600 ppm CDW splitting. This should be readily apparent in our experiment. TaSe$_2$ \cite{suits1980confirmation} and NbSe$_2$ \cite{skripov199577se} similarly exhibit CDW splittings of about 500 ppm for $^{77}$Se, which similar to $^{125}$Te has no nuclear quadrupole splitting, giving further indication of the expected magnitude. Thus, there is no evidence for a CDW in our spectra, even though the reappearance of the Dirac carriers evidenced in our results appears to match the onset of the 3DQHE phase identified in Ref.~\cite{tang2019three}. We speculate that this onset leads instead to an intrinsic 3DQHE, perhaps due to the unusual Dirac-cone dispersion which puts the system closer to a quasi-2D behavior.


In conclusion, we observe a field-induced Dirac gap opening at 150 K in NMR studies of ZrTe$_5$. This occurs when the Dirac electron and hole densities are nearly equal, suggesting an exciton condensate mechanism. The abrupt nature of this transition points to the importance of a normal-electron pocket in stabilizing the condensate. Below 50 K, the gap-opened state dissolves, however we do not see evidence of a CDW state as has been proposed.

\begin{acknowledgments}
This work was supported by Texas A\&M University and the Robert A. Welch Foundation, Grant No. A-1526.
\end{acknowledgments}

\bibliography{ZrTe5_b}

\begin{thebibliography}{39}
\expandafter\ifx\csname natexlab\endcsname\relax\def\natexlab#1{#1}\fi
\expandafter\ifx\csname bibnamefont\endcsname\relax
  \def\bibnamefont#1{#1}\fi
\expandafter\ifx\csname bibfnamefont\endcsname\relax
  \def\bibfnamefont#1{#1}\fi
\expandafter\ifx\csname citenamefont\endcsname\relax
  \def\citenamefont#1{#1}\fi
\expandafter\ifx\csname url\endcsname\relax
  \def\url#1{\texttt{#1}}\fi
\expandafter\ifx\csname urlprefix\endcsname\relax\def\urlprefix{URL }\fi
\providecommand{\bibinfo}[2]{#2}
\providecommand{\eprint}[2][]{\url{#2}}

\bibitem[{\citenamefont{Okada et~al.}(1982)\citenamefont{Okada, Sambongi, Ido,
  Tazuke, Aoki, and Fujita}}]{okada1982negative}
\bibinfo{author}{\bibfnamefont{S.}~\bibnamefont{Okada}},
  \bibinfo{author}{\bibfnamefont{T.}~\bibnamefont{Sambongi}},
  \bibinfo{author}{\bibfnamefont{M.}~\bibnamefont{Ido}},
  \bibinfo{author}{\bibfnamefont{Y.}~\bibnamefont{Tazuke}},
  \bibinfo{author}{\bibfnamefont{R.}~\bibnamefont{Aoki}}, \bibnamefont{and}
  \bibinfo{author}{\bibfnamefont{O.}~\bibnamefont{Fujita}},
  \bibinfo{journal}{J. Phys. Soc. Jpn.} \textbf{\bibinfo{volume}{51}},
  \bibinfo{pages}{460} (\bibinfo{year}{1982}).

\bibitem[{\citenamefont{Wei et~al.}(2012)\citenamefont{Wei, Chao, and
  Aji}}]{wei2012excitonic}
\bibinfo{author}{\bibfnamefont{H.}~\bibnamefont{Wei}},
  \bibinfo{author}{\bibfnamefont{S.-P.} \bibnamefont{Chao}}, \bibnamefont{and}
  \bibinfo{author}{\bibfnamefont{V.}~\bibnamefont{Aji}},
  \bibinfo{journal}{Phys. Rev. Lett.} \textbf{\bibinfo{volume}{109}},
  \bibinfo{pages}{196403} (\bibinfo{year}{2012}).

\bibitem[{\citenamefont{Wang and Zhang}(2013)}]{wang2013chiral}
\bibinfo{author}{\bibfnamefont{Z.}~\bibnamefont{Wang}} \bibnamefont{and}
  \bibinfo{author}{\bibfnamefont{S.-C.} \bibnamefont{Zhang}},
  \bibinfo{journal}{Phys. Rev. B} \textbf{\bibinfo{volume}{87}},
  \bibinfo{pages}{161107} (\bibinfo{year}{2013}).

\bibitem[{\citenamefont{Roy and Sau}(2015)}]{roy2015magnetic}
\bibinfo{author}{\bibfnamefont{B.}~\bibnamefont{Roy}} \bibnamefont{and}
  \bibinfo{author}{\bibfnamefont{J.~D.} \bibnamefont{Sau}},
  \bibinfo{journal}{Phys. Rev. B} \textbf{\bibinfo{volume}{92}},
  \bibinfo{pages}{125141} (\bibinfo{year}{2015}).

\bibitem[{\citenamefont{Liu et~al.}(2016)\citenamefont{Liu, Yuan, Zhang, Jin,
  Narayan, Luo, Chen, Yang, Zou, Wu et~al.}}]{liu2016zeeman}
\bibinfo{author}{\bibfnamefont{Y.}~\bibnamefont{Liu}},
  \bibinfo{author}{\bibfnamefont{X.}~\bibnamefont{Yuan}},
  \bibinfo{author}{\bibfnamefont{C.}~\bibnamefont{Zhang}},
  \bibinfo{author}{\bibfnamefont{Z.}~\bibnamefont{Jin}},
  \bibinfo{author}{\bibfnamefont{A.}~\bibnamefont{Narayan}},
  \bibinfo{author}{\bibfnamefont{C.}~\bibnamefont{Luo}},
  \bibinfo{author}{\bibfnamefont{Z.}~\bibnamefont{Chen}},
  \bibinfo{author}{\bibfnamefont{L.}~\bibnamefont{Yang}},
  \bibinfo{author}{\bibfnamefont{J.}~\bibnamefont{Zou}},
  \bibinfo{author}{\bibfnamefont{X.}~\bibnamefont{Wu}}, \bibnamefont{et~al.},
  \bibinfo{journal}{Nat. Commun.} \textbf{\bibinfo{volume}{7}},
  \bibinfo{pages}{12516} (\bibinfo{year}{2016}).

\bibitem[{\citenamefont{Zhang et~al.}(2016)\citenamefont{Zhang, Hutasoit, Sun,
  Yan, Xu, and Liu}}]{zhang2016topological}
\bibinfo{author}{\bibfnamefont{R.-X.} \bibnamefont{Zhang}},
  \bibinfo{author}{\bibfnamefont{J.~A.} \bibnamefont{Hutasoit}},
  \bibinfo{author}{\bibfnamefont{Y.}~\bibnamefont{Sun}},
  \bibinfo{author}{\bibfnamefont{B.}~\bibnamefont{Yan}},
  \bibinfo{author}{\bibfnamefont{C.}~\bibnamefont{Xu}}, \bibnamefont{and}
  \bibinfo{author}{\bibfnamefont{C.-X.} \bibnamefont{Liu}},
  \bibinfo{journal}{Phys. Rev. B} \textbf{\bibinfo{volume}{93}},
  \bibinfo{pages}{041108} (\bibinfo{year}{2016}).

\bibitem[{\citenamefont{Li et~al.}(2016)\citenamefont{Li, Kharzeev, Zhang,
  Huang, Pletikosi{\'c}, Fedorov, Zhong, Schneeloch, Gu, and
  Valla}}]{li2016chiral}
\bibinfo{author}{\bibfnamefont{Q.}~\bibnamefont{Li}},
  \bibinfo{author}{\bibfnamefont{D.~E.} \bibnamefont{Kharzeev}},
  \bibinfo{author}{\bibfnamefont{C.}~\bibnamefont{Zhang}},
  \bibinfo{author}{\bibfnamefont{Y.}~\bibnamefont{Huang}},
  \bibinfo{author}{\bibfnamefont{I.}~\bibnamefont{Pletikosi{\'c}}},
  \bibinfo{author}{\bibfnamefont{A.~V.} \bibnamefont{Fedorov}},
  \bibinfo{author}{\bibfnamefont{R.~D.} \bibnamefont{Zhong}},
  \bibinfo{author}{\bibfnamefont{J.~A.} \bibnamefont{Schneeloch}},
  \bibinfo{author}{\bibfnamefont{G.~D.} \bibnamefont{Gu}}, \bibnamefont{and}
  \bibinfo{author}{\bibfnamefont{T.}~\bibnamefont{Valla}},
  \bibinfo{journal}{Nat. Phys.} \textbf{\bibinfo{volume}{12}},
  \bibinfo{pages}{550} (\bibinfo{year}{2016}).

\bibitem[{\citenamefont{Tang et~al.}(2019)\citenamefont{Tang, Ren, Wang, Zhong,
  Schneeloch, Yang, Yang, Lee, Gu, Qiao et~al.}}]{tang2019three}
\bibinfo{author}{\bibfnamefont{F.}~\bibnamefont{Tang}},
  \bibinfo{author}{\bibfnamefont{Y.}~\bibnamefont{Ren}},
  \bibinfo{author}{\bibfnamefont{P.}~\bibnamefont{Wang}},
  \bibinfo{author}{\bibfnamefont{R.}~\bibnamefont{Zhong}},
  \bibinfo{author}{\bibfnamefont{J.}~\bibnamefont{Schneeloch}},
  \bibinfo{author}{\bibfnamefont{S.~A.} \bibnamefont{Yang}},
  \bibinfo{author}{\bibfnamefont{K.}~\bibnamefont{Yang}},
  \bibinfo{author}{\bibfnamefont{P.~A.} \bibnamefont{Lee}},
  \bibinfo{author}{\bibfnamefont{G.}~\bibnamefont{Gu}},
  \bibinfo{author}{\bibfnamefont{Z.}~\bibnamefont{Qiao}}, \bibnamefont{et~al.},
  \bibinfo{journal}{Nature} \textbf{\bibinfo{volume}{569}},
  \bibinfo{pages}{537} (\bibinfo{year}{2019}).

\bibitem[{\citenamefont{Zhang et~al.}(2019)\citenamefont{Zhang, Wang, Guo, Zhu,
  Zhang, Yang, Wang, Qu, Pi, Lu et~al.}}]{zhang2019anomalous}
\bibinfo{author}{\bibfnamefont{J.~L.} \bibnamefont{Zhang}},
  \bibinfo{author}{\bibfnamefont{C.~M.} \bibnamefont{Wang}},
  \bibinfo{author}{\bibfnamefont{C.~Y.} \bibnamefont{Guo}},
  \bibinfo{author}{\bibfnamefont{X.~D.} \bibnamefont{Zhu}},
  \bibinfo{author}{\bibfnamefont{Y.}~\bibnamefont{Zhang}},
  \bibinfo{author}{\bibfnamefont{J.~Y.} \bibnamefont{Yang}},
  \bibinfo{author}{\bibfnamefont{Y.~Q.} \bibnamefont{Wang}},
  \bibinfo{author}{\bibfnamefont{Z.}~\bibnamefont{Qu}},
  \bibinfo{author}{\bibfnamefont{L.}~\bibnamefont{Pi}},
  \bibinfo{author}{\bibfnamefont{H.-Z.} \bibnamefont{Lu}},
  \bibnamefont{et~al.}, \bibinfo{journal}{Phys. Rev. Lett.}
  \textbf{\bibinfo{volume}{123}}, \bibinfo{pages}{196602}
  (\bibinfo{year}{2019}).

\bibitem[{\citenamefont{Martino et~al.}(2019)\citenamefont{Martino, Crassee,
  Eguchi, Santos-Cottin, Zhong, Gu, Berger, Rukelj, Orlita, Homes
  et~al.}}]{martino2019two}
\bibinfo{author}{\bibfnamefont{E.}~\bibnamefont{Martino}},
  \bibinfo{author}{\bibfnamefont{I.}~\bibnamefont{Crassee}},
  \bibinfo{author}{\bibfnamefont{G.}~\bibnamefont{Eguchi}},
  \bibinfo{author}{\bibfnamefont{D.}~\bibnamefont{Santos-Cottin}},
  \bibinfo{author}{\bibfnamefont{R.~D.} \bibnamefont{Zhong}},
  \bibinfo{author}{\bibfnamefont{G.~D.} \bibnamefont{Gu}},
  \bibinfo{author}{\bibfnamefont{H.}~\bibnamefont{Berger}},
  \bibinfo{author}{\bibfnamefont{Z.}~\bibnamefont{Rukelj}},
  \bibinfo{author}{\bibfnamefont{M.}~\bibnamefont{Orlita}},
  \bibinfo{author}{\bibfnamefont{C.~C.} \bibnamefont{Homes}},
  \bibnamefont{et~al.}, \bibinfo{journal}{Phys. Rev. Lett.}
  \textbf{\bibinfo{volume}{122}}, \bibinfo{pages}{217402}
  (\bibinfo{year}{2019}).

\bibitem[{\citenamefont{Jiang et~al.}(2020)\citenamefont{Jiang, Wang, Zhao,
  Dun, Huang, Wu, Mourigal, Zhou, Pan, Ozerov et~al.}}]{jiang2020unraveling}
\bibinfo{author}{\bibfnamefont{Y.}~\bibnamefont{Jiang}},
  \bibinfo{author}{\bibfnamefont{J.}~\bibnamefont{Wang}},
  \bibinfo{author}{\bibfnamefont{T.}~\bibnamefont{Zhao}},
  \bibinfo{author}{\bibfnamefont{Z.~L.} \bibnamefont{Dun}},
  \bibinfo{author}{\bibfnamefont{Q.}~\bibnamefont{Huang}},
  \bibinfo{author}{\bibfnamefont{X.~S.} \bibnamefont{Wu}},
  \bibinfo{author}{\bibfnamefont{M.}~\bibnamefont{Mourigal}},
  \bibinfo{author}{\bibfnamefont{H.~D.} \bibnamefont{Zhou}},
  \bibinfo{author}{\bibfnamefont{W.}~\bibnamefont{Pan}},
  \bibinfo{author}{\bibfnamefont{M.}~\bibnamefont{Ozerov}},
  \bibnamefont{et~al.}, \bibinfo{journal}{Phys. Rev. Lett.}
  \textbf{\bibinfo{volume}{125}}, \bibinfo{pages}{046403}
  (\bibinfo{year}{2020}).

\bibitem[{\citenamefont{J{\'e}rome et~al.}(1967)\citenamefont{J{\'e}rome, Rice,
  and Kohn}}]{jerome1967excitonic}
\bibinfo{author}{\bibfnamefont{D.}~\bibnamefont{J{\'e}rome}},
  \bibinfo{author}{\bibfnamefont{T.~M.} \bibnamefont{Rice}}, \bibnamefont{and}
  \bibinfo{author}{\bibfnamefont{W.}~\bibnamefont{Kohn}},
  \bibinfo{journal}{Phys. Rev.} \textbf{\bibinfo{volume}{158}},
  \bibinfo{pages}{462} (\bibinfo{year}{1967}).

\bibitem[{\citenamefont{Kotov et~al.}(2012)\citenamefont{Kotov, Uchoa, Pereira,
  Guinea, and Neto}}]{kotov2012electron}
\bibinfo{author}{\bibfnamefont{V.~N.} \bibnamefont{Kotov}},
  \bibinfo{author}{\bibfnamefont{B.}~\bibnamefont{Uchoa}},
  \bibinfo{author}{\bibfnamefont{V.~M.} \bibnamefont{Pereira}},
  \bibinfo{author}{\bibfnamefont{F.}~\bibnamefont{Guinea}}, \bibnamefont{and}
  \bibinfo{author}{\bibfnamefont{A.~C.} \bibnamefont{Neto}},
  \bibinfo{journal}{Rev. Mod. Phys.} \textbf{\bibinfo{volume}{84}},
  \bibinfo{pages}{1067} (\bibinfo{year}{2012}).

\bibitem[{\citenamefont{Scherer et~al.}(2018)\citenamefont{Scherer, Honerkamp,
  Rudenko, Stepanov, Lichtenstein, and Katsnelson}}]{scherer2018excitonic}
\bibinfo{author}{\bibfnamefont{M.~M.} \bibnamefont{Scherer}},
  \bibinfo{author}{\bibfnamefont{C.}~\bibnamefont{Honerkamp}},
  \bibinfo{author}{\bibfnamefont{A.~N.} \bibnamefont{Rudenko}},
  \bibinfo{author}{\bibfnamefont{E.~A.} \bibnamefont{Stepanov}},
  \bibinfo{author}{\bibfnamefont{A.~I.} \bibnamefont{Lichtenstein}},
  \bibnamefont{and} \bibinfo{author}{\bibfnamefont{M.~I.}
  \bibnamefont{Katsnelson}}, \bibinfo{journal}{Phys. Rev. B}
  \textbf{\bibinfo{volume}{98}}, \bibinfo{pages}{241112}
  (\bibinfo{year}{2018}).

\bibitem[{\citenamefont{Rudenko et~al.}(2018)\citenamefont{Rudenko, Stepanov,
  Lichtenstein, and Katsnelson}}]{rudenko2018excitonic}
\bibinfo{author}{\bibfnamefont{A.~N.} \bibnamefont{Rudenko}},
  \bibinfo{author}{\bibfnamefont{E.~A.} \bibnamefont{Stepanov}},
  \bibinfo{author}{\bibfnamefont{A.~I.} \bibnamefont{Lichtenstein}},
  \bibnamefont{and} \bibinfo{author}{\bibfnamefont{M.~I.}
  \bibnamefont{Katsnelson}}, \bibinfo{journal}{Phys. Rev. Lett.}
  \textbf{\bibinfo{volume}{120}}, \bibinfo{pages}{216401}
  (\bibinfo{year}{2018}).

\bibitem[{\citenamefont{Wang et~al.}(2020{\natexlab{a}})\citenamefont{Wang,
  Liu, Wan, and Zhang}}]{wang2020quantum}
\bibinfo{author}{\bibfnamefont{J.-R.} \bibnamefont{Wang}},
  \bibinfo{author}{\bibfnamefont{G.-Z.} \bibnamefont{Liu}},
  \bibinfo{author}{\bibfnamefont{X.}~\bibnamefont{Wan}}, \bibnamefont{and}
  \bibinfo{author}{\bibfnamefont{C.}~\bibnamefont{Zhang}},
  \bibinfo{journal}{Phys. Rev. B} \textbf{\bibinfo{volume}{101}},
  \bibinfo{pages}{245151} (\bibinfo{year}{2020}{\natexlab{a}}).

\bibitem[{\citenamefont{Tian et~al.}(2019)\citenamefont{Tian, Ghassemi, and
  Ross}}]{tian2019dirac}
\bibinfo{author}{\bibfnamefont{Y.}~\bibnamefont{Tian}},
  \bibinfo{author}{\bibfnamefont{N.}~\bibnamefont{Ghassemi}}, \bibnamefont{and}
  \bibinfo{author}{\bibfnamefont{J.~H.} \bibnamefont{Ross},
  \bibfnamefont{Jr.}}, \bibinfo{journal}{Phys. Rev. B}
  \textbf{\bibinfo{volume}{100}}, \bibinfo{pages}{165149}
  (\bibinfo{year}{2019}).

\bibitem[{\citenamefont{Shahi et~al.}(2018)\citenamefont{Shahi, Singh, Sun,
  Zhao, Chen, Lv, Li, Yan, Mandrus, and Cheng}}]{shahi2018bipolar}
\bibinfo{author}{\bibfnamefont{P.}~\bibnamefont{Shahi}},
  \bibinfo{author}{\bibfnamefont{D.~J.} \bibnamefont{Singh}},
  \bibinfo{author}{\bibfnamefont{J.~P.} \bibnamefont{Sun}},
  \bibinfo{author}{\bibfnamefont{L.~X.} \bibnamefont{Zhao}},
  \bibinfo{author}{\bibfnamefont{G.~F.} \bibnamefont{Chen}},
  \bibinfo{author}{\bibfnamefont{Y.~Y.} \bibnamefont{Lv}},
  \bibinfo{author}{\bibfnamefont{J.}~\bibnamefont{Li}},
  \bibinfo{author}{\bibfnamefont{J.-Q.} \bibnamefont{Yan}},
  \bibinfo{author}{\bibfnamefont{D.~G.} \bibnamefont{Mandrus}},
  \bibnamefont{and} \bibinfo{author}{\bibfnamefont{J.-G.} \bibnamefont{Cheng}},
  \bibinfo{journal}{Phys. Rev. X} \textbf{\bibinfo{volume}{8}},
  \bibinfo{pages}{021055} (\bibinfo{year}{2018}).

\bibitem[{\citenamefont{Inamo}(1996)}]{inamo1996125te}
\bibinfo{author}{\bibfnamefont{M.}~\bibnamefont{Inamo}},
  \bibinfo{journal}{Chem. Lett.} \textbf{\bibinfo{volume}{25}},
  \bibinfo{pages}{17} (\bibinfo{year}{1996}).

\bibitem[{\citenamefont{Weng et~al.}(2014)\citenamefont{Weng, Dai, and
  Fang}}]{weng2014transition}
\bibinfo{author}{\bibfnamefont{H.}~\bibnamefont{Weng}},
  \bibinfo{author}{\bibfnamefont{X.}~\bibnamefont{Dai}}, \bibnamefont{and}
  \bibinfo{author}{\bibfnamefont{Z.}~\bibnamefont{Fang}},
  \bibinfo{journal}{Phys. Rev. X} \textbf{\bibinfo{volume}{4}},
  \bibinfo{pages}{011002} (\bibinfo{year}{2014}).

\bibitem[{\citenamefont{Okv{\'a}tovity
  et~al.}(2019)\citenamefont{Okv{\'a}tovity, Yasuoka, Baenitz, Simon, and
  D{\'o}ra}}]{okvatovity2019nuclear}
\bibinfo{author}{\bibfnamefont{Z.}~\bibnamefont{Okv{\'a}tovity}},
  \bibinfo{author}{\bibfnamefont{H.}~\bibnamefont{Yasuoka}},
  \bibinfo{author}{\bibfnamefont{M.}~\bibnamefont{Baenitz}},
  \bibinfo{author}{\bibfnamefont{F.}~\bibnamefont{Simon}}, \bibnamefont{and}
  \bibinfo{author}{\bibfnamefont{B.}~\bibnamefont{D{\'o}ra}},
  \bibinfo{journal}{Phys. Rev. B} \textbf{\bibinfo{volume}{99}},
  \bibinfo{pages}{115107} (\bibinfo{year}{2019}).

\bibitem[{\citenamefont{Maebashi et~al.}(2019)\citenamefont{Maebashi, Hirosawa,
  Ogata, and Fukuyama}}]{maebashi2019nuclear}
\bibinfo{author}{\bibfnamefont{H.}~\bibnamefont{Maebashi}},
  \bibinfo{author}{\bibfnamefont{T.}~\bibnamefont{Hirosawa}},
  \bibinfo{author}{\bibfnamefont{M.}~\bibnamefont{Ogata}}, \bibnamefont{and}
  \bibinfo{author}{\bibfnamefont{H.}~\bibnamefont{Fukuyama}},
  \bibinfo{journal}{J. Phys. Chem. Solids} \textbf{\bibinfo{volume}{128}},
  \bibinfo{pages}{138} (\bibinfo{year}{2019}).

\bibitem[{\citenamefont{Tian et~al.}(2018)\citenamefont{Tian, Zhu, Ren,
  Ghassemi, Conant, Wang, Ren, and Ross}}]{tian2018native}
\bibinfo{author}{\bibfnamefont{Y.}~\bibnamefont{Tian}},
  \bibinfo{author}{\bibfnamefont{H.}~\bibnamefont{Zhu}},
  \bibinfo{author}{\bibfnamefont{W.}~\bibnamefont{Ren}},
  \bibinfo{author}{\bibfnamefont{N.}~\bibnamefont{Ghassemi}},
  \bibinfo{author}{\bibfnamefont{E.}~\bibnamefont{Conant}},
  \bibinfo{author}{\bibfnamefont{Z.}~\bibnamefont{Wang}},
  \bibinfo{author}{\bibfnamefont{Z.}~\bibnamefont{Ren}}, \bibnamefont{and}
  \bibinfo{author}{\bibfnamefont{J.~H.} \bibnamefont{Ross},
  \bibfnamefont{Jr.}}, \bibinfo{journal}{Phys. Chem. Chem. Phys.}
  \textbf{\bibinfo{volume}{20}}, \bibinfo{pages}{21960} (\bibinfo{year}{2018}).

\bibitem[{\citenamefont{Tian et~al.}(2020)\citenamefont{Tian, Ghassemi, and
  Ross}}]{tian2020topological}
\bibinfo{author}{\bibfnamefont{Y.}~\bibnamefont{Tian}},
  \bibinfo{author}{\bibfnamefont{N.}~\bibnamefont{Ghassemi}}, \bibnamefont{and}
  \bibinfo{author}{\bibfnamefont{J.~H.} \bibnamefont{Ross},
  \bibfnamefont{Jr.}}, \bibinfo{journal}{Phys. Rev. B}
  \textbf{\bibinfo{volume}{102}}, \bibinfo{pages}{165149}
  (\bibinfo{year}{2020}).

\bibitem[{\citenamefont{Tay et~al.}(2020)\citenamefont{Tay, Shang, Puphal,
  Pomjakushina, Ott, and Shiroka}}]{tay2020unusual}
\bibinfo{author}{\bibfnamefont{D.}~\bibnamefont{Tay}},
  \bibinfo{author}{\bibfnamefont{T.}~\bibnamefont{Shang}},
  \bibinfo{author}{\bibfnamefont{P.}~\bibnamefont{Puphal}},
  \bibinfo{author}{\bibfnamefont{E.}~\bibnamefont{Pomjakushina}},
  \bibinfo{author}{\bibfnamefont{H.-R.} \bibnamefont{Ott}}, \bibnamefont{and}
  \bibinfo{author}{\bibfnamefont{T.}~\bibnamefont{Shiroka}},
  \bibinfo{journal}{{Phys. Rev. B}} \textbf{\bibinfo{volume}{102}},
  \bibinfo{pages}{241109} (\bibinfo{year}{2020}).

\bibitem[{\citenamefont{Wang et~al.}(2020{\natexlab{b}})\citenamefont{Wang,
  Honjo, Zhao, Chen, Matano, Zhou, and Zheng}}]{wang2020landau}
\bibinfo{author}{\bibfnamefont{C.~G.} \bibnamefont{Wang}},
  \bibinfo{author}{\bibfnamefont{Y.}~\bibnamefont{Honjo}},
  \bibinfo{author}{\bibfnamefont{L.~X.} \bibnamefont{Zhao}},
  \bibinfo{author}{\bibfnamefont{G.~F.} \bibnamefont{Chen}},
  \bibinfo{author}{\bibfnamefont{K.}~\bibnamefont{Matano}},
  \bibinfo{author}{\bibfnamefont{R.}~\bibnamefont{Zhou}}, \bibnamefont{and}
  \bibinfo{author}{\bibfnamefont{G.-q.} \bibnamefont{Zheng}},
  \bibinfo{journal}{Phys. Rev. B} \textbf{\bibinfo{volume}{101}},
  \bibinfo{pages}{241110} (\bibinfo{year}{2020}{\natexlab{b}}).

\bibitem[{\citenamefont{Morice et~al.}(2020)\citenamefont{Morice, Lettl, Kopp,
  and Kampf}}]{morice2020optical}
\bibinfo{author}{\bibfnamefont{C.}~\bibnamefont{Morice}},
  \bibinfo{author}{\bibfnamefont{E.}~\bibnamefont{Lettl}},
  \bibinfo{author}{\bibfnamefont{T.}~\bibnamefont{Kopp}}, \bibnamefont{and}
  \bibinfo{author}{\bibfnamefont{A.~P.} \bibnamefont{Kampf}},
  \bibinfo{journal}{Phys. Rev. B} \textbf{\bibinfo{volume}{102}},
  \bibinfo{pages}{155138} (\bibinfo{year}{2020}).

\bibitem[{\citenamefont{Chen et~al.}(2015)\citenamefont{Chen, Chen, Song,
  Schneeloch, Gu, Wang, and Wang}}]{chen2015magnetoinfrared}
\bibinfo{author}{\bibfnamefont{R.~Y.} \bibnamefont{Chen}},
  \bibinfo{author}{\bibfnamefont{Z.~G.} \bibnamefont{Chen}},
  \bibinfo{author}{\bibfnamefont{X.-Y.} \bibnamefont{Song}},
  \bibinfo{author}{\bibfnamefont{J.~A.} \bibnamefont{Schneeloch}},
  \bibinfo{author}{\bibfnamefont{G.~D.} \bibnamefont{Gu}},
  \bibinfo{author}{\bibfnamefont{F.}~\bibnamefont{Wang}}, \bibnamefont{and}
  \bibinfo{author}{\bibfnamefont{N.~L.} \bibnamefont{Wang}},
  \bibinfo{journal}{Phys. Rev. Lett.} \textbf{\bibinfo{volume}{115}},
  \bibinfo{pages}{176404} (\bibinfo{year}{2015}).

\bibitem[{\citenamefont{Sun et~al.}(2020)\citenamefont{Sun, Cao, Cui, Zhu, Ma,
  Wang, Zhuo, Cheng, Wang, Wan et~al.}}]{sun2020large}
\bibinfo{author}{\bibfnamefont{Z.}~\bibnamefont{Sun}},
  \bibinfo{author}{\bibfnamefont{Z.}~\bibnamefont{Cao}},
  \bibinfo{author}{\bibfnamefont{J.}~\bibnamefont{Cui}},
  \bibinfo{author}{\bibfnamefont{C.}~\bibnamefont{Zhu}},
  \bibinfo{author}{\bibfnamefont{D.}~\bibnamefont{Ma}},
  \bibinfo{author}{\bibfnamefont{H.}~\bibnamefont{Wang}},
  \bibinfo{author}{\bibfnamefont{W.}~\bibnamefont{Zhuo}},
  \bibinfo{author}{\bibfnamefont{Z.}~\bibnamefont{Cheng}},
  \bibinfo{author}{\bibfnamefont{Z.}~\bibnamefont{Wang}},
  \bibinfo{author}{\bibfnamefont{X.}~\bibnamefont{Wan}}, \bibnamefont{et~al.},
  \bibinfo{journal}{npj Quantum Mater.} \textbf{\bibinfo{volume}{5}},
  \bibinfo{pages}{36} (\bibinfo{year}{2020}).

\bibitem[{sm()}]{sm}
\bibinfo{note}{See Supplemental Material at [URL will be inserted by publisher]
  for simulation results, which includes
  Refs.~\cite{tang2019three,chen2015magnetoinfrared,liu2016zeeman,sun2020large}}.

\bibitem[{\citenamefont{Carter et~al.}(1977)\citenamefont{Carter, Bennett, and
  Kahan}}]{carter1977metallic}
\bibinfo{author}{\bibfnamefont{G.~C.} \bibnamefont{Carter}},
  \bibinfo{author}{\bibfnamefont{L.~H.} \bibnamefont{Bennett}},
  \bibnamefont{and} \bibinfo{author}{\bibfnamefont{D.~J.} \bibnamefont{Kahan}},
  \emph{\bibinfo{title}{{Metallic shifts in NMR: a review of the theory and
  comprehensive critical data compilation of metallic materials}}}
  (\bibinfo{publisher}{Pergamon, New York}, \bibinfo{year}{1977}).

\bibitem[{\citenamefont{Burkov et~al.}(2011)\citenamefont{Burkov, Hook, and
  Balents}}]{burkov2011topological}
\bibinfo{author}{\bibfnamefont{A.~A.} \bibnamefont{Burkov}},
  \bibinfo{author}{\bibfnamefont{M.~D.} \bibnamefont{Hook}}, \bibnamefont{and}
  \bibinfo{author}{\bibfnamefont{L.}~\bibnamefont{Balents}},
  \bibinfo{journal}{Phys. Rev. B} \textbf{\bibinfo{volume}{84}},
  \bibinfo{pages}{235126} (\bibinfo{year}{2011}).

\bibitem[{\citenamefont{Janssen and Herbut}(2016)}]{janssen2016excitonic}
\bibinfo{author}{\bibfnamefont{L.}~\bibnamefont{Janssen}} \bibnamefont{and}
  \bibinfo{author}{\bibfnamefont{I.~F.} \bibnamefont{Herbut}},
  \bibinfo{journal}{Phys. Rev. B} \textbf{\bibinfo{volume}{93}},
  \bibinfo{pages}{165109} (\bibinfo{year}{2016}).

\bibitem[{\citenamefont{Wang et~al.}(2017)\citenamefont{Wang, Liu, and
  Zhang}}]{wang2017excitonic}
\bibinfo{author}{\bibfnamefont{J.-R.} \bibnamefont{Wang}},
  \bibinfo{author}{\bibfnamefont{G.-Z.} \bibnamefont{Liu}}, \bibnamefont{and}
  \bibinfo{author}{\bibfnamefont{C.-J.} \bibnamefont{Zhang}},
  \bibinfo{journal}{Phys. Rev. B} \textbf{\bibinfo{volume}{95}},
  \bibinfo{pages}{075129} (\bibinfo{year}{2017}).

\bibitem[{\citenamefont{Pan and Shindou}(2019)}]{pan2019ground}
\bibinfo{author}{\bibfnamefont{Z.}~\bibnamefont{Pan}} \bibnamefont{and}
  \bibinfo{author}{\bibfnamefont{R.}~\bibnamefont{Shindou}},
  \bibinfo{journal}{Phys. Rev. B} \textbf{\bibinfo{volume}{100}},
  \bibinfo{pages}{165124} (\bibinfo{year}{2019}).

\bibitem[{\citenamefont{Fjellv{\aa}g and
  Kjekshus}(1986)}]{fjellvaag1986structural}
\bibinfo{author}{\bibfnamefont{H.}~\bibnamefont{Fjellv{\aa}g}}
  \bibnamefont{and} \bibinfo{author}{\bibfnamefont{A.}~\bibnamefont{Kjekshus}},
  \bibinfo{journal}{Solid State Commun.} \textbf{\bibinfo{volume}{60}},
  \bibinfo{pages}{91} (\bibinfo{year}{1986}).

\bibitem[{\citenamefont{Ross and Slichter}(1992)}]{ross1992nuclear}
\bibinfo{author}{\bibfnamefont{J.~H.} \bibnamefont{Ross}, \bibfnamefont{Jr.}}
  \bibnamefont{and} \bibinfo{author}{\bibfnamefont{C.~P.}
  \bibnamefont{Slichter}}, in \emph{\bibinfo{booktitle}{Nuclear Spectroscopy on
  Charge Density Wave Systems}} (\bibinfo{publisher}{Springer},
  \bibinfo{year}{1992}), pp. \bibinfo{pages}{113--175}.

\bibitem[{\citenamefont{Suits et~al.}(1980)\citenamefont{Suits, Couturie, and
  Slichter}}]{suits1980confirmation}
\bibinfo{author}{\bibfnamefont{B.~H.} \bibnamefont{Suits}},
  \bibinfo{author}{\bibfnamefont{S.}~\bibnamefont{Couturie}}, \bibnamefont{and}
  \bibinfo{author}{\bibfnamefont{C.~P.} \bibnamefont{Slichter}},
  \bibinfo{journal}{Phys. Rev. Lett.} \textbf{\bibinfo{volume}{45}},
  \bibinfo{pages}{194} (\bibinfo{year}{1980}).

\bibitem[{\citenamefont{Skripov et~al.}(1995)\citenamefont{Skripov, Sibirtsev,
  Cherepanov, and Aleksashin}}]{skripov199577se}
\bibinfo{author}{\bibfnamefont{A.~V.} \bibnamefont{Skripov}},
  \bibinfo{author}{\bibfnamefont{D.~S.} \bibnamefont{Sibirtsev}},
  \bibinfo{author}{\bibfnamefont{Y.~G.} \bibnamefont{Cherepanov}},
  \bibnamefont{and} \bibinfo{author}{\bibfnamefont{B.~A.}
  \bibnamefont{Aleksashin}}, \bibinfo{journal}{J. Phys. Condens. Matter}
  \textbf{\bibinfo{volume}{7}}, \bibinfo{pages}{4479} (\bibinfo{year}{1995}).

\end{thebibliography}

\end{document}